\documentclass[a4paper]{spie}  %>>> use for US letter paper
\usepackage[]{graphicx}

\title{Sparse aperture differential piston measurements using the pyramid wave-front sensor} 

\author{Arcidiacono Carmelo\supit{a}, Chen  Xinyang\supit{b}
,Yan Zhaojun\supit{b}
, Zheng Lixin\supit{b}
, Agapito Guido\supit{c}
, Wang Chaoyan\supit{b}
, Zhu Nenghong\supit{b}
, Zhu Liyun\supit{b}
, Cai Jianqing\supit{b}
and Tang Zhenghong\supit{b}
\skiplinehalf
\supit{a}INAF - Osservatorio Astronomico di Bologna, Via Ranzani 1, I-40127 Bologna, Italy; \\
\supit{b}SHAO - Shanghai Astronomical Observatory, Nandan Road, 80, 200030 Shanghai, China;\\
\supit{c}INAF - Osservatorio Astrofisico di Arcetri, Largo Enrico Fermi 5, I-50125 Firenze, Italy;
}

\authorinfo{Further author information: (Send correspondence to Carmelo Arcidiacono)\\C.A.: E-mail: carmelo.arcidiacono@oabo.inaf.it, Telephone:  +39 051 2095 704}
\begin{document} 
  \maketitle 

%%%%%%%%%%%%%%%%%%%%%%%%%%%%%%%%%%%%%%%%%%%%%%%%%%%%%%%%%%%%% 
\begin{abstract}
In this paper we report on the laboratory experiment we settled in the Shanghai Astronomical Observatory (SHAO) to investigate the pyramid wave-front sensor (WFS) ability to measure the differential piston on a sparse aperture. The ultimate goal is to verify the ability of the pyramid WFS work in closed loop to perform the phasing of the primary mirrors of a sparse Fizeau imaging telescope. In the experiment we installed on the optical bench we performed various test checking the ability to flat the wave-front using a deformable mirror and to measure the signal of the differential piston on a two pupils setup. These steps represent the background from which we start to perform full closed loop operation on multiple apertures. These steps were also useful to characterize the achromatic double pyramids (double prisms) manufactured in the SHAO optical workshop.
\end{abstract}

%>>>> Include a list of keywords after the abstract 


%%%%%%%%%%%%%%%%%%%%%%%%%%%%%%%%%%%%%%%%%%%%%%%%%%%%%%%%%%%%%
\section{INTRODUCTION}
\label{sec:intro}  % \label{} allows reference to this section
Existing large telescope, such as the 10m Keck\cite{1990SPIE.1236..996J}, and foreseen extremely large Telescope, the Giant Magellan Telescope\cite{2012SPIE.8444E..1HJ}, have
a segmented primary mirror. This choice is typically driven by technological challenges and budget considerations that support this solution with respect to a bulky or active primary.\\
The adaptive optics (AO) corrects wave-front aberrations all over the telescope aperture, but to effectively restore images at the diffraction limited resolution the co-phasing of the composing segments is required. From the AO point of view, the pupil segmentation may represent a challenge since the accuracy should go at the level of a small fraction of the observing
wavelength.\\ Moreover
a typical wave-front sensor (WFS) is, in a first instance, sensitive to the first derivative of the wave-front. Geometrical optics returns that Pyramid\cite{pyramid} WFS, Shack Hartmann\cite{hartmann} WFS  and curvature sensor are not sensitive to the phase discontinuities.\\
But interference patterns actually modulate the light close to the edges of the segments with intensities dependent on the phase jumps between those.\\

At the Keck telescopes, several techniques for the co-phasing of the primary segments have been proposed\cite{1994SPIE.2199..622C,2000ApOpt..39.4706C} and tested. The Active Phasing Experiment\cite{2005SPIE.5894..306G} (APE) of the European Southern Observatory (ESO) compared different concepts of phasing sensors. These tests were performed in the laboratory and on sky at the Very Large Telescope (VLT). The sensors considered were: 
curvature sensor, the pyramid sensor\cite{2008SPIE.7012E..3DP} (PYPS), the Shack-Hartmann type phasing sensor\cite{2008SPIE.7012E..3AM} (SHAPS) and the
Zernike Phase Contrast Sensor\cite{2006SPIE.6267E..34D} (ZEUS).\\ The Shack-Hartmann is the sensor actually used at Keck\cite{2000SPIE.4003..188C} telescopes. In this article, we report about the activities performed in the Shanghai Astronomical Observatory (SHAO) using the pyramid WFS as part of a general study about sparse aperture interferometry.\\
A point should be raised about segments phasing when the separation of the segment become large:
larger the segment separation higher is the modulation frequency of the interference pattern. This effect makes the differential piston measurements more challenging in the case of large separations of the segments, such as in the Large Binocular Telescope\cite{2010ApOpt..49..115H} (LBT) or Giant Magellan Telescope (GMT) cases.\\
\section{Pyramid Wave-front Sensor}
Pyramid Wavefront sensor is particularly suitable for the co-phasing. The numerical simulations performed in 2001\cite{2002ESOC...58..161E} and later detailed for the Large Binocular Telescope case in Verinaud~\&~Esposito (2002)~\cite{2002OptL...27..470V,2002ESOC...58..153V} show the ability of the pyramid WFS to sense differential piston on sparse apertures. The light close to the edge of the segmented pupil is modulated by a diffraction pattern depending on the differential piston of the segments. More exactly the light interference between discontinued sub-pupil portions produces a signal in the position of the image of the pupil corresponding to the loci of the phase step. Another effects is that a modulation intensity pattern, that is dependent on the sine of the phase step value, is superimposed over the pupil illumination. This statement was initially demonstrated on a laboratory setup\cite{2003SPIE.5169...72E} and eventually on sky\cite{2009Msngr.136...25G}.
\section{Experimental Setup}
\begin{figure}
   %\begin{center}
   \begin{tabular}{l}
   \includegraphics[width=14cm]{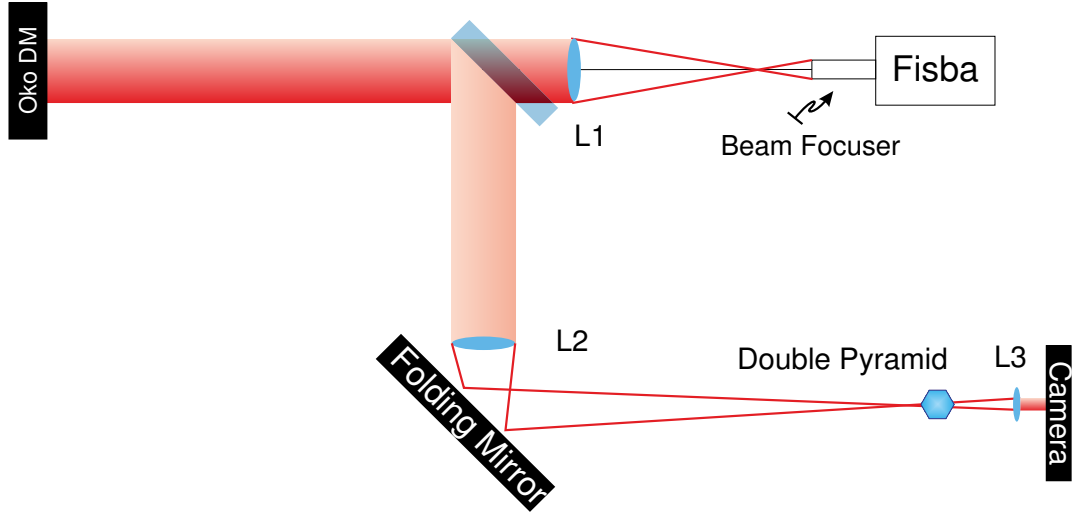}\\
   \end{tabular}
  % \end{center}
   \caption[example] 
%>>>> use \label inside caption to get Fig. number with \ref{}
   { \label{fig:scheme} 
The scheme of the optical setup used for testing the pyramid WFS ability to sense the differential piston over sparse aperture.}
   \end{figure}

\begin{figure}
   \begin{center}
   \begin{tabular}{c}
   \includegraphics[height=7cm]{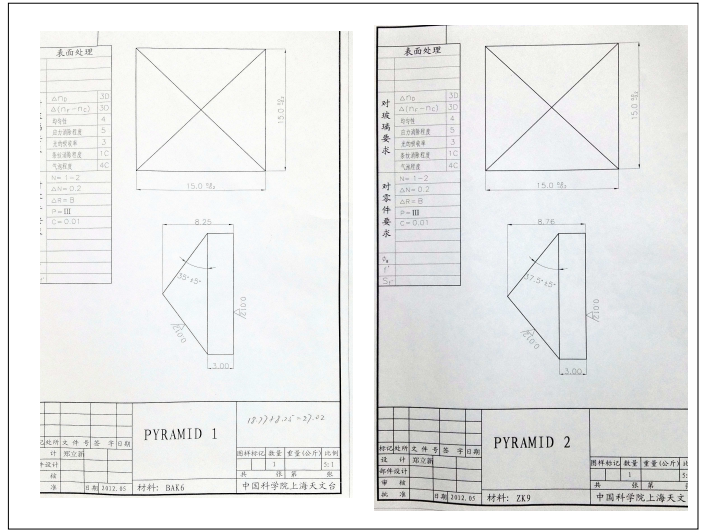}\\
   \end{tabular}
   \end{center}
   \caption[example] 
%>>>> use \label inside caption to get Fig. number with \ref{}
   { \label{fig:pyr} 
Pyramids technical specifications used for the manufacture.}
   \end{figure}

We aligned on an optical bench an experimental setup devoted to the determination of the pyramid WFS ability to measure the differential piston between two different sub--pupils, acting as a sparse aperture. Using a deformable mirror (DM) to control the differential piston term and $\mu$Phase\textsuperscript{\textregistered}~500 interferometer from Trioptics that we used to cross check the effectiveness of the piston actuation. In Figure~\ref{fig:scheme} a sketch of the optical design. The DM was an OKO\textsuperscript{\textregistered} Tech 109-channels
piezoelectric. This DM has a clear aperture of 53mm and it is packed on
on a box of $100\times100\times60 mm$. The mirror figure is controlled by 109 piezoelectric
actuators positioned on an orthogonal grid. The actuators pitch is of 4.3mm.
The pyramid we used was manufactured at the optical workshop of the SHAO and actually consists of an achromatic system since it is the result of the assembly
of two pyramidic prisms on a base to base configuration. Details are in Table~\ref{tab:pyr} and Figure~\ref{fig:pyr}. 
%-------------
%-------------
Referring to the scheme in Figure~\ref{fig:scheme} :
\begin{itemize}
\item ``Fisba" is the interferometer which propagates the light of a $632.8nm$ He-Ne laser source;
\item L1 is a lens aligned in order to have its focus corresponding to the one of an objective (Beam focuser in the figure) inserted on the collimated beam at the exit of the interferometer: a collimated beam of the size of the L1 diameter is propagated in the forward direction and along the optical axis;
\item a 50/50 beam splitter allows the refraction of a transmitted beam back towards the interferometer and the reflection of the image of the DM pupil towards L2;
\item L2 is a lens which focal length is 1000mm and at this focal distance has on the of the 50/50 beam splitter the deformable mirror and on the other the (first) pin of the double pyramid;
\item L3 is a lens which focal length is  50mm, it works as objective and has on a side the (first) pyramid pin on the other the camera detector both at the focal distance.
\end{itemize}
%-------------
   \begin{figure}
   \begin{center}
   \begin{tabular}{c}
   \includegraphics[width=16cm]{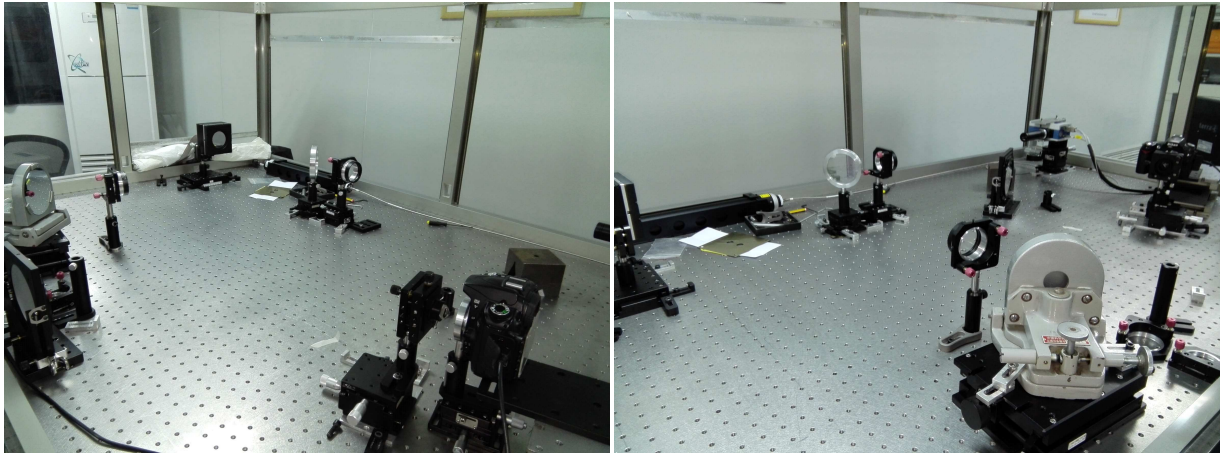}\\
   \end{tabular}
   \end{center}
   \caption[example] 
%>>>> use \label inside caption to get Fig. number with \ref{}
   { \label{fig:pic} 
Two pictures taken from different points of view. Images show the optical setup as it was realized in the laboratory.}
   \end{figure} 
In the two pictures in Figure~\ref{fig:pic} the instrumental setup is well visible.  
Finally the pupil can be set by using a mask in front of the DM.\\

For the selection of the lenses above we considered that the four-pupils image projected on the camera must be:
\begin{itemize}
\item large/small enough to be sampled by the camera
\item Smaller than the separation of the pupil
\item L3 focal length should allow to have the pupil well separated once the detector of the camera is conjugated at the pupil plane.
\end{itemize}

The detector used was a Nikon\textsuperscript{\textregistered} D90\textsuperscript{\textregistered}: it is a 12.3-megapixel digital single-lens 
reflex camera with detector dimension of $23.6mm \times 15.8mm $ with
$ 4288 \times 2848$ pixels of $5.5~micron$ pixelsize.
The divergence angle of the pyramid, as from Tozzi et al 2008\cite{doublepyr}, is:
%\begin{equation}
\newline
\begin{center}
$\theta = \left(n_{1} - 1\right) \cdot \alpha_{1} - \left(n_{2} - 1\right) \cdot  \alpha_{2}$\\
\end{center}

%\end{equation}
 using $\alpha_1=35\deg$ $\alpha_2=37\deg 30'$,
$n_1=1.563, n_2=1.638$ we have back a divergence angle $\theta=4\deg  12'$.
In table~\ref{tab:pyr}  and in Figure~\ref{fig:pyr} the characteristics of the pyramid, see also Arcidiacono~(2005)\cite{2005OptCo.252..239A} for reference.
%-------------
%-------------

\begin{figure}
   \begin{center}
   \begin{tabular}{c}
   \includegraphics[width=12cm]{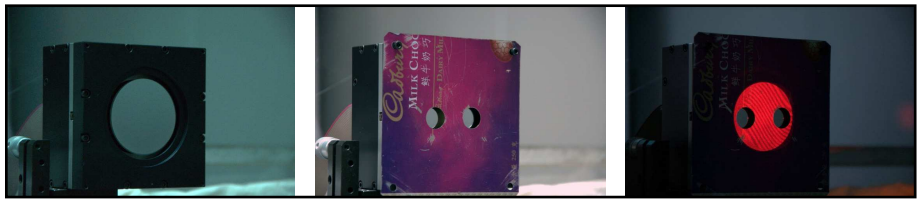}\\
   \end{tabular}
   \end{center}
   \caption[example] 
%>>>> use \label inside caption to get Fig. number with \ref{}
   { \label{fig:oko} 
These pictures show the Deformable Mirror without and with the sub-pupil mask installed. On the right picture, the beam from the interferometer illuminates the mask.}
   \end{figure}

\begin{figure}
   \begin{center}
   \begin{tabular}{c}
   \includegraphics[height=7cm]{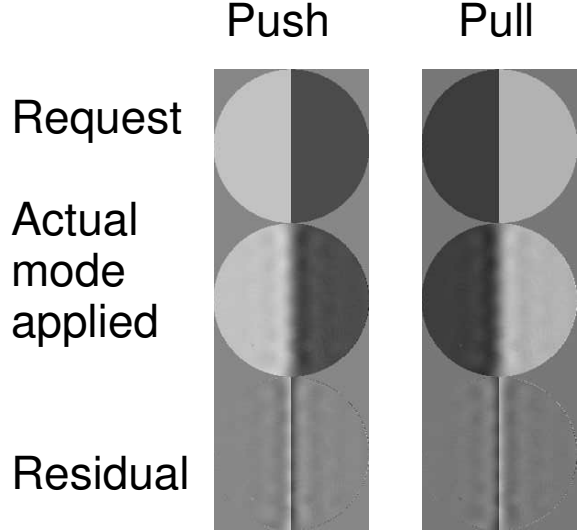}\\
   \end{tabular}
   \end{center}
   \caption[example] 
%>>>> use \label inside caption to get Fig. number with \ref{}
   { \label{fig:pushpull} 
The picture shows the push and pull actuation of the piston term on the left and right column respectively. From the top to the bottom the requested phase shape, the mode applied (as composed through the measured DM influence functions) and finally the phase residual (the difference requested - IF fit).}
   \end{figure}

%% This table is carefully placed in the source file to make 
%% it appear at bottom of page, but above the footnotes.  
%% Use of [h] in following command forces table to appear "here".
\begin{table}[h]
\caption{Pyramids Specifications} 
\label{tab:pyr}
\begin{center}       
\begin{tabular}{|l|l|} %% this creates two columns
%% |l|l| to left justify each column entry
%% |c|c| to center each column entry
%% use of \rule[]{}{} below opens up each row
\hline
\rule[-1ex]{0pt}{3.5ex}  Pyramid 1 &  \\
\hline
\rule[-1ex]{0pt}{3.5ex}  Refractive Index & 1.5  \\
\hline
\rule[-1ex]{0pt}{3.5ex}  Angle & 35degrees  \\
\hline
\rule[-1ex]{0pt}{3.5ex}  Pyramid 2  & \\
\hline
\rule[-1ex]{0pt}{3.5ex}  Refractive Index & 1.638  \\
\hline
\rule[-1ex]{0pt}{3.5ex}  Angle & 37.5degrees   \\
\hline 
\end{tabular}
\end{center}
\end{table} 
Eventually the pupil images have a diameter, $s$ of :
\begin{equation}
s = l \frac{f_3}{f_2}\\
\end{equation}
\begin{equation}
s = 50 \frac{50}{1000} = 2.5mm = 450px\\
\end{equation}
given that $l$ is the aperture size, $f_2$ and $f_3$ the focal length of L2 and L3.
For the four-pupils separation, $d$ we got:
\begin{equation}
d = 2\theta f_3\\
\end{equation}
\begin{equation}
d = 2 \cdot 0.073rad \cdot 50mm = 7.35mm = 1335px\\
\end{equation}
The deformable mirror has a clear diameter of 53mm, slightly larger than beam collimated by the 2~inches L1 lens. The dimension of the unmasked beam size was even smaller, about 50mm, because of the L1 lens holder vignetting.
%-------------
   \begin{figure}
   \begin{center}
   \begin{tabular}{c}
   \includegraphics[height=7cm]{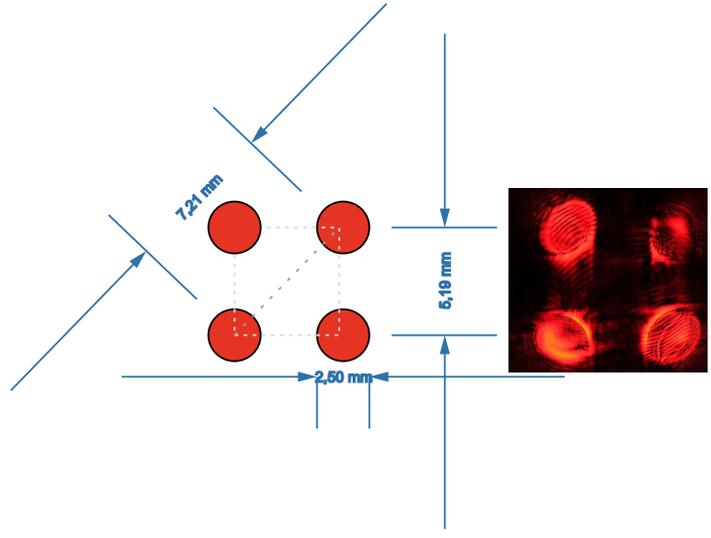}\\
   \end{tabular}
   \end{center}
   \caption[example] 
%>>>> use \label inside caption to get Fig. number with \ref{}
   { \label{fig:pyrpup} 
On the left the expected geometry of the four-pupils image and on the right a picture of what was actually obtained.}
   \end{figure} 
We placed in front of the DM a pupil mask cutting two identical sub-pupils well spaced. The dimension of the pupils and
separation were defined to allow the injection of a pure differential piston between those by pushing the left half of the actuators and pulling the other side (or the opposite way to have the opposite differential piston). The projections of the two sub-pupils on the DM were separated by 7 actuators,
reducing the cross-talk of the piston term with other optical aberrations. The final mirror figure representing the differential piston was computed by projecting the differential piston mode on the space of the DM optical influence function, see figure~\ref{fig:pushpull} and~\ref{fig:oko} and the following section.

\section{Deformable Mirror Calibration}
\begin{figure}
   \begin{center}
   \begin{tabular}{c}
   \includegraphics[width=12cm]{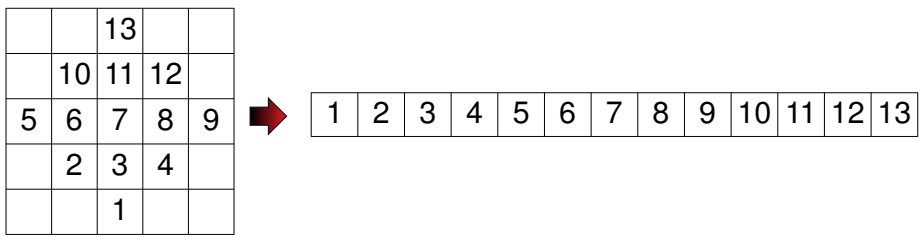}\\
   \end{tabular}
   \end{center}
   \caption[example] 
%>>>> use \label inside caption to get Fig. number with \ref{}
   { \label{fig:arrvec} 
We show the way we map an array onto a vector.}
   \end{figure}

To properly control the optical figure of the DM we had to know the influence function (the optical response of the each actuator the DM) of the commands sent by the user through the computer. \\
Typically we refer to the command as the command vector, $V$, that controls the actuation of the 109 channels of the deformable mirror.  \\
Since that the DM has a linear response to the actuation of the commands we can build an operation matrix, friendly called Voltage2Fisba, $V2F$, that projects a command vector for the DM
into an optical response as measured by the interferometer, $WF$. 
\begin{equation}
WF = V2F \times V
\end{equation}
$WF$ is the measurement performed by the interferometer which was transformed from a 2D image onto a 1D vector. See figure~\ref{fig:arrvec},

The simplest way to build the $V2F$ is to directly measure it using the interferometer. We follow a push-pull strategy to obtain the influence function of each actuator, $IF_i$, from two consecutive wave-front measurements corresponding to the application, with opposite voltages, $v$,  of the actuator of interest. We measured the influence function as the optical response of the i-actuator. 
We computed the $IF_i$ through the equation~\ref{eq:pp}: defining as $A_{i,v}$ and $A_{i,-v}$
the optical response measured by the interferometer as the response to the application of the voltages $v$ and $-v$ respectively. We may write it as
\begin{equation}
IF_i = \left( A_{i,v} - A_{i,-v} \right) / \left(2 \cdot v\right)
\label{eq:pp}
\end{equation}
In this way the $V2F$ array is measured, having on the column the $IF_i$ array, written as a column vector as long as the number of pixels $npx$ composing the illuminated part of the pupil as imaged on the interferometer:

\begin{equation}
	V2F = \left(\begin{array}[h]{c c c}
	IF_{1,1} & \cdots & IF_{1,109} \\
	\vdots   & \ddots & \vdots \\
	IF_{npx,1} & \cdots & IF_{npx,109} \\
	\end{array} \right)
	\end{equation}

%-------------

The $V2F$ matrix can be inverted through (for example) a Singular Value Decomposition and computing
the inverse $F2V=V2F^{-1}$, which projects an interferometer phase measurement, still $WF$, into a vector of commands, still $V$,
for the deformable mirror.
\begin{equation}
V  = F2V \times WF 
\end{equation}
\begin{equation}
WF = V2F \times V
\end{equation}
%-------------
   \begin{figure}
   \begin{center}
   \begin{tabular}{c}
   \includegraphics[width=16cm]{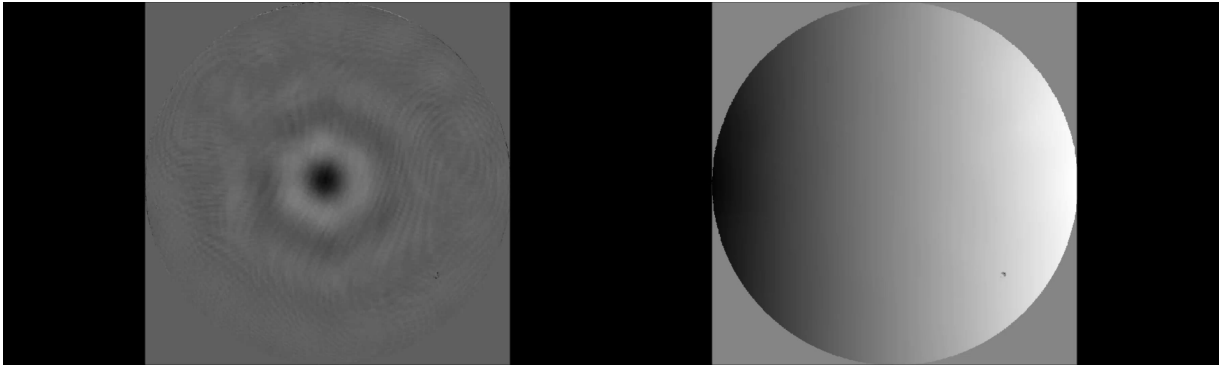}\\
   \end{tabular}
   \end{center}
   \caption[example] 
%>>>> use \label inside caption to get Fig. number with \ref{}
   { \label{fig:1act1zer} 
The optical response measured by the interferometer for the central actuator and the first Zernike mode, respectively on the left and the right.}
   \end{figure} 

The $F2V$ obtained in this way was used to build a matrix of DM commands corresponding to the Zernike Polynomials base. The Zernike modes were initially analytically
computed in the space of the WF as measured by the interferometer. Here we refer to the base as $ZNK$:
\begin{equation}
V_{i,ZNK} = F2V \times ZNK_i
\end{equation}
The $V_{i,ZNK}$ commands, which inject on the DM the i-th Zernike polynomial shape ($ZNK_i$), is the least mean square solution of the linear system:
\begin{equation}
ZNK_i = V2F \times V_{i,ZNK} 
\end{equation}
More generally for all the Zernike mode of interest  (we computed 140 Zernike) we may write the projection on the command space of the Zernike polynomials base:
\begin{equation}
V_{ZNK} = F2V \times ZNK
\end{equation}

We repeated the push-pull procedure applying the $V_{i,ZNK}$ commands. We used the modal base of the Zernike modes to measure a more reliable matrix
for the projection of the commands space into the interferometer measurement one, $V2F'$.
We may write the push-pull sequence of commands applying the Zernike polynomial
in a way similar to the equation~\ref{eq:pp}: 
\begin{equation}
WF_{i-ZNK} = \left(A_{i-ZNK,v}-A_{i-ZNK,-v}\right)/\left(2v\right)
\end{equation}
where $A_{i-ZNK,v}$ is the optical response measured by the interferometer to the command 
$V_{i,ZNK}$ corresponding to the i-th Zernike. A this point a new $V2F$, call it $V2F'$, array may be computed from the $WF_{ZNK}$ array,
as well as its inverse, $F2V'$:
\[
V2F' = V_{ZNK}^{-1} \times WF_{ZNK}
\]
The new Zernike polynomial base $V'_{ZNK}=F2V'\times ZNK$ projected on the command space produces an optical response, $A_{ZNK}$, which is better matching the desired analytical model, $ZNK$:
\begin{itemize}
\item $A_{ZNK} - ZNK = $ has an average residual of 75nm RMS on average for the Focus and astigmatism considering 1micron applied,
\item $A'_{ZNK} - ZNK = $ has an average residual of 5nm RMS, as above.
\end{itemize}

In order to speed up the measurements collection we developed a software tool (a GUI) which was installed on the computer used for the interferometer control and that was able to send to the DM the series of push-pull commands for each actuator. After the initial single actuator sequence, it was upgraded to send also the series push and pull actuation of the Zernike modes.
The final $F2V’$ matrix was used to compute the command vectors corresponding to push and pull of the piston on half of the full pupil in order to apply a piston term on the single sub-pupils.
%-------------
\section{Piston Footprint}
The goal is to have evidence of a signal coming from the insertion of a differential piston 
on the two sub-pupils. 

Initial Deformable Mirror shape with command vector $V$ = [0,\ldots,0], had an aberration of 365nm rms,
to be compared the measurement done with a flat mirror inserted just in front of the DM (and following L1) for the calibration that returned~165nm rms.

After the second iteration of measurement and correction using still the $F2V’$ we got a residual 110nm rms.

On the best available pattern, we add the push (positive)  and pull (negative) differential piston command retrieving that the two halves of the mirror were actually flat,
within the error of the measurements (about 3nm rms).
Of course, we also measured the phase on the junction of the two side that presented a clear slope.
Once installed the mask with the two sub-pupils we were ready with pyramid measurement.

\begin{figure}
   \begin{center}
   \begin{tabular}{c}
   \includegraphics[height=5cm]{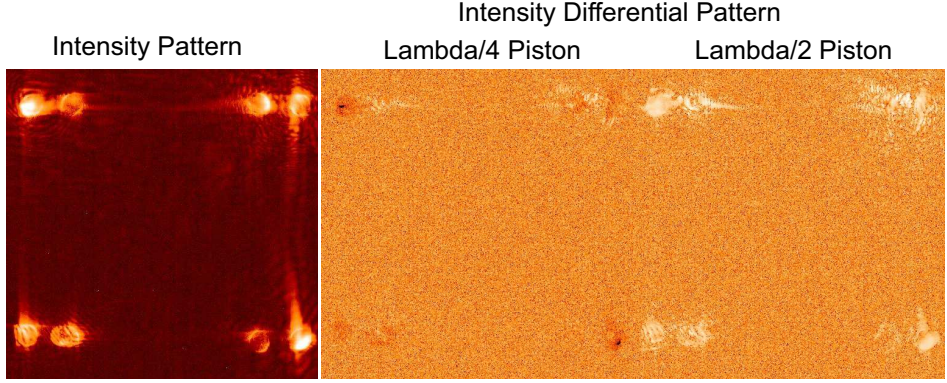}\\
   \end{tabular}
   \end{center}
   \caption[example] 
%>>>> use \label inside caption to get Fig. number with \ref{}
   { \label{fig:intensity} 
In the left part of the figure the intensity pattern as measured on the camera detector. On the right the push-pull difference of the application of two differential piston of a quarter of lambda and of half of a lambda.}
   \end{figure}

Unfortunately, we have suffered for the effect of some fringing introduced most probably by the 50/50 beam splitter and visible on the detector. The beam splitter produces a back reflected pupil image which is visible on the intensity pattern on the left side of the Figure~\ref{fig:intensity}. Moreover, since the pyramid is not modulated and the focal plane image of the source on the pyramid pin is quasi diffraction-limited the pyramid sensitivity to aberration is very high. A relatively small residual (the 110nm above) produces a pretty important variation of the illumination of the pupil and some pupil distortion. In the following month, we succeeded to obtain a better alignment at the level of $30 to 35nm$ rms.
Here we report on the first test performed before this update of the alignment.
On the aligned setup we perform push pull series of differential piston mode with different amplitude. In particular the case of a quarter of lambda and of half a lambda are shown in figure~\ref{fig:intensity}. The signal relative to the differential piston stays in the fringes visible in within the two sub-pupils.
\section{Conclusion}
We did not perform a quantitative analysis on this data since we have succeeded to reach a much better alignment (as we actually did, but unfortunately not in time to be shown here) and the pupil distortions were too big.
However, looking into the frames we got, we catch the signal from the differential piston, which is visible on the fringes within the two sub-pupils. We plan to collect a new, and undistorted, data set to be processed in order to give quantitative results. Further progress in the direction of the use of the pyramid sensor as differential piston sensor is on-going in parallel to closed loop test for the co-phase of sparse aperture sub-pupils in broad band visible light.

We have shown that pyramid prism manufactured in SHAO provided a good optical quality and that pyramid pin is sharp enough to let the pyramid WFS working even in the diffraction limited regime without any modulation of the light.

%%%%%%%%%%%%%%%%%%%%%%%%%%%%%%%%%%%%%%%%%%%%%%%%%%%%%%%%%%%%%
%%%%%%%%%%%%%%%%%%%%%%%%%%%%%%%%%%%%%%%%%%%%%%%%%%%%%%%%%%%%%
\acknowledgments     %>>>> equivalent to \section*{ACKNOWLEDGMENTS}       
The first author wishes to thank Chinese Academy of Science (CAS) for the important opportunity to visit SHAO and the surrounding astronomical facilities. In particular, he wishes to thank SHAO and the interferometry group for the warm hospitality and namely the Academic Prof. Zhu. The study was supported by the CAS President's International Scientist Initiative (PIFI) 2015.

%%%%%%%%%%%%%%%%%%%%%%%%\neq%%%%%%%%%%%%%%%%%%%%%%%%%%%%%%%%%%%%%
%%%%% References %%%%%

\bibliography{report}   %>>>> bibliography data in report.bib
\bibliographystyle{spiebib}   %>>>> makes bibtex use spiebib.bst

\end{document}